# Tutorial: The Ubiquitous Skiplist, its Variants, and Applications in Modern Big Data Systems

Venkata Sai Pavan Kumar Vadrevu, Lu Xing, Walid G. Aref
Purdue University
West Lafayette, Indiana, , USA
{vvadrevu,xingl,aref}@purdue.edu

## ABSTRACT

The Skiplist, or skip list, originally designed as an in-memory data structure, has attracted a lot of attention in recent years as a main-memory component in many NoSQL, cloud-based, and big data systems. Unlike the B-tree, the skiplist does not need complex rebalancing mechanisms, but it still shows expected logarithmic performance. It supports a variety of operations, including insert, point read, and range queries. To make the skiplist more versatile, many optimizations have been applied to its node structure, construction algorithm, list structure, concurrent access, to name a few. Many variants of the skiplist have been proposed and experimented with, in many big-data system scenarios.

In addition to being a main-memory component, the skiplist also serves as a core index in systems to address problems including write amplification, write stalls, sorting, range query processing, etc. In this tutorial, we present a comprehensive overview of the skiplist, its variants, optimizations, and various use cases of how big data and NoSQL systems make use of skiplists. Throughout this tutorial, we demonstrate the advantages of using a skiplist or skiplist-like structures in modern data systems.

## 1 INTRODUCTION

The skip list, first introduced by Pugh [52], is becoming very popular, and a new word "skiplist" has been invented for that. It is a probabilistic alternative to balanced trees [52], but requires less effort to balance. Many systems require a fast and concurrent main-memory component that maintains sorted order over the incoming data and skiplist is the ideal candidate for these systems. Skiplist has been utilized in many data systems, e.g., [1–4, 10, 28, 35, 70] as the only or partial memory component for fast concurrent insertions and later being merged with other sorted parts. Skiplist or skiplist-like structures are also used as standalone indexes in systems, e.g., [34, 38, 40, 55].

**Tutorial Overview.** We plan for a 1.5-hr tutorial that is split into 5 sections.

**(1) Skiplist Basics (10 min).** We begin by introducing the original design of the skiplist by Pugh [52, 54]. We dissect its structure in detail, and hint at potential usages and optimizations.

**(2) Skiplist and Tree Equivalence (10 min).** We highlight the duality between skiplists and tree structures. Interestingly, this duality gives rise to mapping of concurrency control algorithms among the two structures.

**(3) Variants of Skiplists (20 min).** We overview a variety of optimizations that apply to skiplists, including adaptivity, determinism, supporting concurrency, node structure; and numerous skiplist variants, e.g., skiplists over DRAM, SSD, and Intel Optane Persistent Memory, NUMA-aware skiplists, and skiplists optimized for certain data types and certain operations.

**(4) Skiplists in Modern Data Systems (40 min).** We overview how various data systems put skiplists into action in modern data systems, especially in main memory components and buffers of write-optimized data systems. We give an overview of the data systems centered around skiplists. We highlight skiplist features that make these systems outperform others under certain workloads.

**(5) Other Use Cases of Skiplists (10 min).** Skiplists are widely used in other systems areas, e.g., in network protocols, distributed systems [16, 43, 49, 58], video streaming [62], and blockchain [57]. We will highlight other systems structures that are built using skiplists, e.g., the skip graphs [7]. We conclude by discussing open problems and challenges in future research.

**Target Audience.** We target a broad range of audience including students, database researchers and practitioners who seek to acquire the basic structure and evolution of skiplists, its potential optimizations, how skiplists can benefit modern data systems, and key features of skiplist-based data stores.

**Related Tutorials.** This is the first tutorial on skiplists and their use in modern data systems. Recently, other tutorials have addressed trending indexes, e.g., read-optimizations for LSM-trees [56] and indexing on Persistent Memory [36].

## 2 TUTORIAL OUTLINE

**2.1. Basic Skiplist.** The Skiplist is a probabilistic data structure introduced by Pugh [52]. The expected cost of both insert and point read is $O(\log n)$ where $n$ is the number of items in the skiplist. Basically, the skiplist is a hierarchy of linked lists, where the data in the upper-Level-$(i + 1)$ list is a subset of the data in the linked list in Level-$i$. Search starts at the topmost coarse grained level where unnecessary comparisons are skipped and moves downwards to a finer search in the lower levels. The height of a node is chosen randomly by a predefined probability $p$ that determines the fraction of nodes to be promoted to the level above. Algorithms have been developed to construct optimal or near-optimal skiplists [50].

**2.2. Duality of Skiplists and Balanced Trees.**
In the first skiplist paper, Pugh mentions that both balanced trees and skiplists have performance bounds of the same order but skiplists are easier to implement than balanced trees [52]. [47] has shown a correspondence between 2-3 (2-3-4) trees and 1-2 (1-2-3) deterministic skiplists. [41] maps deterministic skiplists to B-trees by adding two B-tree structures. A skip tree [45] is an intermediate structure between skiplists and B-trees, and [20] describes how to interpret a skiplist as a randomly-balanced binary search tree.

**2.3 Skiplist Concurrency Control**

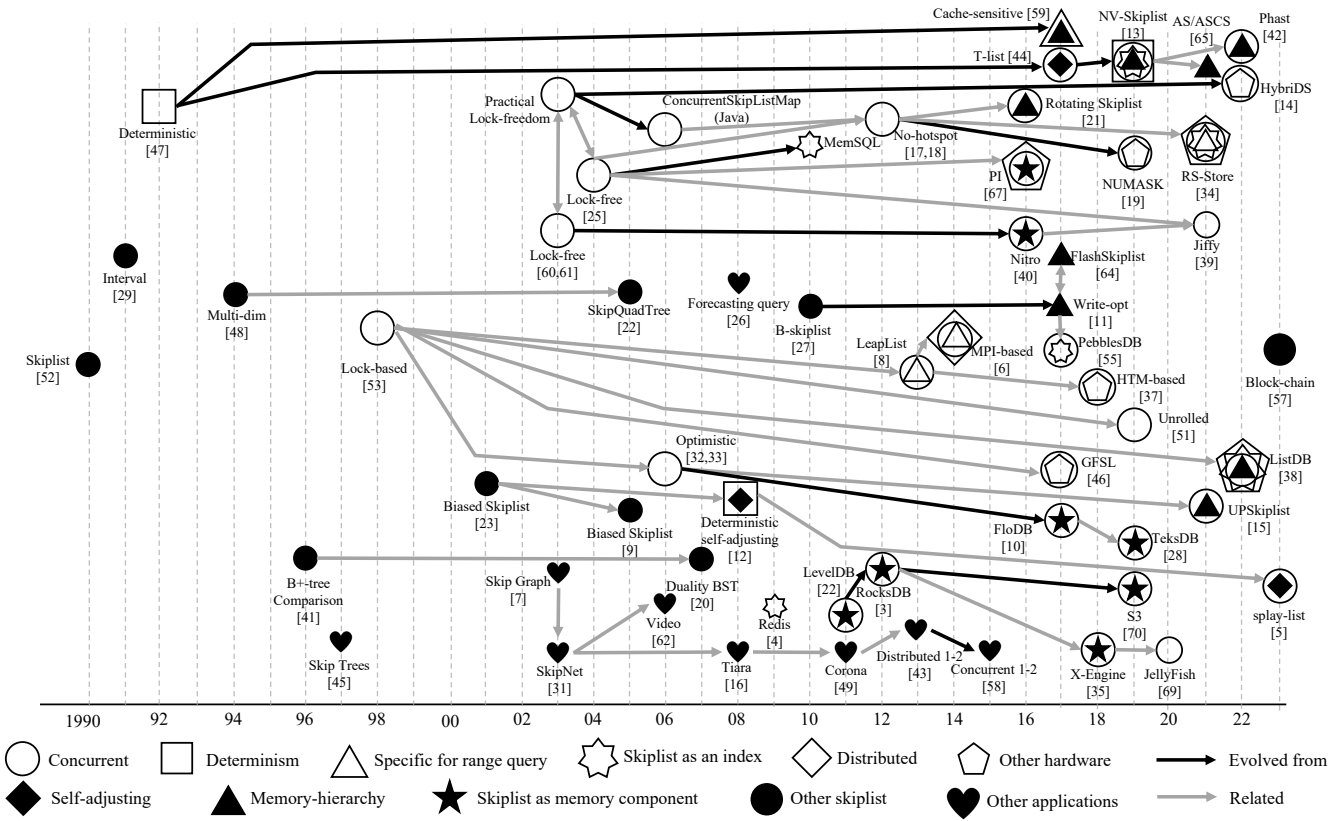

Figure 1: Skiplist variants and their use cases. Notice that since all skiplist variants evolve from the original 1990 Skiplist, edges from it to all others are omitted from the figure for better visualization.

**Lock-based.** Pugh discusses concurrent skiplists [53] that use write locks when inserting or deleting an item. Herlihy et al. [32, 33] propose an optimistic lock-based skiplist that requires short lock-based validation before insert and delete. Unrolled skiplists [51] group multiple keys per node, and use advanced locking primitive based on group mutual exclusion to achieve concurrency.

**Non-blocking.** A lock-free skiplist uses a Compare-And-Swap primitive [25, 60, 61]. As the higher level is accessed more often and is prone to contention, [17, 18] decouples each update into an eager abstract modification for fast commit, and a selective lazy adaptation to apply changes in the background. The rotating skiplist combines a lock-free rotation of a tree with the skiplist [21].

**Transactional Memory.** Transactional memory is another way to achieve synchronization. A Hardware Transactional Memory (HTM)-based skiplist is presented in [37].

**Multi-Version Concurrency Control (MVCC).** MVCC applies over skiplists. The skiplist of both X-Engine [35] and JellyFish [69] organize a vertical list to store multiple versions of the same key. Jiffy [39] maintains a vertical list of versions that store all data in the range between two neighbor nodes in the bottom linked list.

### 2.4. Skiplist Variants

**Deterministic, Adaptive, and Biased Skiplists.** In the worst-case, probabilistic skiplists may have unbounded performance. In contrast, a deterministic skiplist has a worst-case bounded cost. The 1-2/1-2-3 skiplists [47] are deterministic, and have up to 2-3 elements at Height $h-1$ between any two nodes of Height $h > 1$. [12]'s deterministic skiplist is self-adjusting. The T-list [44] builds a skiplist top-down as a byproduct during search. When a predefined count of pointer chases is reached, a node is promoted to facilitate the next search. In contrast, the splay-list [5] records the number of accesses of each node, and promotes or demotes a node accordingly.

In contrast to assuming uniformity when accessing skiplist elements, biased skiplists [9, 23] handle non-uniform access patterns.

**Skiplists over New Hardware.** Skiplists are designed for various storage media, e.g., DRAM, SSD, and Intel Optane Persistent Memory. Skiplists are adapted for Non-Uniform Memory Access architectures (NUMA), as well as GPU and Near-memory-processing.

**Various Storage Media.** Skiplists are adapted for block devices, SSDs, and Persistent Memory (PM). Write-optimized skiplists [11] allocate buffers within each node. One node contains several pivots that point to child level nodes containing the same copies of the pivots. FlashSkipList [24, 64] associates a buffer with a level, and is later pushed down to lower levels when certain criterion is met. The NV-skiplist [13] is designed for PM, and adopts several designs to reduce the number of PM writes. The UPSkiplist [15] uses epochs to solve the crash-consistent issue in PM. Atomic Skiplists (AS) and Atomic and Selective Consistency Skiplists (ASCS) [65] are write-optimized for PM. PhaST [42] is another skiplist for PM that supports concurrent access and is log-free.

**NUMA-awareness.** Non-Uniform Memory Access (NUMA)-aware skiplists, e.g., PI/PSL [66–68] and Braided Skiplist [38] partition the skiplist into ranges to assign to different NUMA nodes. NUMASK [19] places the upper levels in different NUMA nodes.

**GPU, RDMA and Near-memory-processing.** GFSL [46] is a GPU-based skiplist. R-skiplist [34] is a skiplist over RDMA. A hybrid skiplist [14] is for near-memory processing (NMP).

**Other Skiplist Variants.** There are more optimizations on range querying over skiplists [6, 8, 59] utilizing cache-sensitivity, MPI or software transactional memory. The skiplist can index other data types, e.g., multidimensional data [22, 48] and intervals [29, 30]. The skiplist is also used in processing forecasting queries [26].

## 2.5. Skiplist Applications in Big Data Systems.

**Skiplist as a Memory Component.** Typically, a disk-based key-value store has memory buffers to facilitate data writes. Memory buffers need to be highly concurrent and easy to remain sorted, which makes skiplist a good choice for memory buffers. Many key-value stores use Log-Structured Merge Tree (LSMT) that has a memory component. LevelDB [2] uses skiplists to organize data in-memory. RocksDB's memory buffer is a skiplist [3], and has a HashSkipList variant that links hash buckets in a skiplist. This variant performs less comparisons for point reads but incurs more overhead to keep data sorted for range queries. X-engine [35] supports a multi-version skiplist. Redis [4] uses skiplist in its SortedSet to improve search. HBase [1] also uses skiplists for in-memory data.

**Skiplist + X.** For LSM key-value stores to scale with the size of main memory, auxiliary structures are introduced to work along with skiplists [10, 28, 70]. S3 [70] has two-layered memory. The bottom layer is a semi-sorted skiplist; with unsorted keys but the overall ranges of all nodes are ordered. To improve search, guards in the top layer are chosen by a neural model to skip traversing skiplist nodes. A cache-sensitive index organizes the guards on the top layer. FloDB [10] has two-level memory: a small-sized concurrent hash table at the top and a large-sized concurrent skiplist at the bottom, which favors parallel scans and writes and low-latency updates. The main memory in TeksDB [28] has two complementary structures: a cuckoo hash and a skiplist for fast point reads and range queries.

**Skiplist-based Systems.** Besides being a memory component, a skiplist is also widely used as an index. Nitro [40] uses a lock-free skiplist as its core index. Other components, e.g., MVCC, garbage collection, snapshot manager and recovery manager are added on top. There are other systems that rely on key features of skiplists, e.g., being balanced, support for in-place inserts, keeping data sorted, and supporting range queries.

**Balanced Structure.** One drawback of LSMT is its high write amplification. Data may be written multiple times before reaching the bottom level, which wastes write I/O bandwidth. To alleviate this issue, LSMT groups data into ranges as in B/B+-trees. However, maintaining a tree structure is non-trivial. PebblesDB [55] combines skiplists with LSMT in the fragmented LSMT (FLSM). Each FLSM level uses guards that are chosen from the inserted keys to split that level's key space. Each guard has associated sstables. Guards are chosen from the inserted keys similar to the skiplist.

**In-place Sorted Inserts.** LSMT may experience write stalls. Data is accumulated in DRAM before being flushed to disk. If sorting and compaction are slow and the write buffer cannot be flushed as needed, the incoming writes may be blocked causing high latency. The skiplist handles this issue by maintaining sorted order while allowing fast in-place inserts. ListDB [38] has a NUMA-aware skiplist on persistent memory to realize a skiplist-based key-value store. ListDB is a 3-level LSMT with one volatile level in DRAM; Levels L0 and L1 in PM. ListDB reduces write amplification by eliminating writes upon MemTable flush and merge-sort of persistent data. When inserting data into DRAM's MemTable, same data with (meta-data) is inserted as a log in PM. This log entry with preallocated skiplist pointers is later inserted as a skiplist node into L0. At merge-sort of L0-L1, data is merged in-place with only modifications to skiplist pointers.

**Range Query Processing.** Hash-based key-value stores over RDMA mostly support point queries. A Skiplist, supporting range queries and being highly concurrent, is a good fit for an RDMA-based key-value store. RS-store [34] is a skiplist-based key-value store using RDMA. Its R-skiplist groups multiple entries into one block. The R-skiplist is partitioned to increase scalability with one thread exclusively per partition. If the client-issued request cannot be handled due to high server CPU utilization, the client's request accesses the R-skiplist via RDMA that does not involve the use of server's CPU. This remote access is read-only and the remote range scan does not involve heavy use of server CPU as server returns once with the starting address of the block and client can issue remote RDMA reads to retrieve following blocks.

**Other Applications** Skiplists are heavily used in peer-to-peer video streaming [62, 63], network protocols, history-independent file systems [27], distributed systems [16, 31, 43, 49, 58]. The skiplist-based skip graph [7] is widely used in distributed environments. Skiplist is also being used in blockchain [57].

## 3 BIOGRAPHIES

**VS Pavan Kumar Vadrevu** is a Master's student in Computer Science at Purdue University. His interests lie in the areas of big data systems, databases, and artificial intelligence.

**Lu Xing** is a PhD student in Computer Science at Purdue University advised by Walid G. Aref. Her research interests lie in index optimization and adaptivity.

**Walid G. Aref** is a professor of Computer Science at Purdue. His research interests are in spatial, spatio-temporal, graph, and sensor data management. He is the Editor-in-Chief of the ACM Transactions of Spatial Algorithms and Systems (ACM TSAS).

## ACKNOWLEDGMENTS

Walid G. Aref acknowledges the support of the National Science Foundation under Grant Number IIS-1910216.